\documentclass[12pt,a4paper]{article}
\usepackage[a4paper, total={7in, 9.5in}]{geometry}

\usepackage{amsthm,amsmath,amssymb,amsfonts,bm}
\usepackage{graphicx}
\usepackage{placeins}
\usepackage{epstopdf}
\usepackage{multirow}
\usepackage{hyperref}
\usepackage[labelfont=bf]{caption}
\usepackage{array}
\usepackage{threeparttable}
\usepackage{booktabs}

\def\hang{\hangindent\parindent}
\def\rf{\par\noindent\hang}

\begin{document}

\baselineskip=20pt

\begin{center}
{\bf \Large Confounding-adjustment methods for the causal difference in medians}
\end{center}

\medskip

\begin{center}
{\bf Daisy A. Shepherd*$^{1,2}$, Benjamin R. Baer$^{3}$ \& Margarita Moreno-Betancur$^{1,2}$}
\end{center}

\medskip

\noindent {\sl \rf \small $^{1}$Clinical Epidemiology \& Biostatistics Unit, Department of Paediatrics, The University of Melbourne, VIC, Australia \\
$^{2}$Clinical Epidemiology \& Biostatistics Unit, The Murdoch Children's Research Institute, The Royal Children's Hospital, VIC, Australia \\
$^{3}$Department of Biostatistics and Computational Biology, The University of Rochester, New York, USA}

\bigskip

\noindent {\rf \textbf{Correspondence}: {\bf*}Daisy A. Shepherd, The Murdoch Children's Research Institute, Royal Children's Hospital, VIC, Australia. Email: daisy.shepherd@mcri.edu.au}

\bigskip

\noindent {\rf \textbf{Keywords}: causal inference, skewed outcomes, potential outcomes, difference in medians, confounding, quantile regression, inverse probability weighted, propensity scores, g-computation}

\bigskip

\noindent {\rf \textbf{Abbreviations}: LSAC, The Longitudinal Study of Australian Children; SDQ, Strengths and Difficulties Questionnaire; QR, quantile regression; IP, inverse probability; IPW, inverse probability weighted; PS, propensity score}

\newpage

\begin{center}
{\bf ABSTRACT}
\end{center}
With continuous outcomes, the average causal effect is typically defined using a contrast of expected potential outcomes. However, in the presence of skewed outcome data, the expectation may no longer be meaningful. In practice the typical approach is to either ``ignore or transform” - ignore the skewness altogether or transform the outcome to obtain a more symmetric distribution, although neither approach is entirely satisfactory. Alternatively the causal effect can be redefined as a contrast of median potential outcomes, yet discussion of confounding-adjustment methods to estimate this parameter is limited. In this study we described and compared confounding-adjustment methods to address this gap. The methods considered were multivariable quantile regression, an inverse probability weighted (IPW) estimator, weighted quantile regression and two little-known implementations of g-computation for this problem. Motivated by a cohort investigation in the Longitudinal Study of Australian Children, we conducted a simulation study that found the IPW estimator, weighted quantile regression and g-computation implementations minimised bias when the relevant models were correctly specified, with g-computation additionally minimising the variance. These methods provide appealing alternatives to the common ``ignore or transform” approach and multivariable quantile regression, enhancing our capability to obtain meaningful causal effect estimates with skewed outcome data.


\newpage

\section*{Introduction} \label{sec:intro}

Causal inference is a central goal of health research, aiming to assess how intervening on a given exposure impacts an outcome of interest.\cite{Hernan2020} In a perfect randomized controlled trial (i.e., with no loss to follow-up) causal effects can in principle be directly estimated by comparing the average outcome in those randomised to each exposure level. However, in observational studies estimation of causal effects requires more sophisticated methods, in particular to adjust for potential confounding due to the lack of randomisation.

With continuous outcomes, the average causal effect is typically defined as a contrast of the expected potential outcome under exposure versus under no exposure (i.e., the estimand in the hypothetical target trial we seek to emulate with observational data\cite{Hernan2016, Moreno2021}). However, epidemiological studies may suffer from skewed outcome data, for which the expectation may no longer be interpretable as the central value of the distribution. Examples of skewed outcomes are abundant in health research, particularly in areas using scale scores for measurement (e.g., self-reported quality of life via the PedsQL\cite{Varni1999}, childhood behaviour via the SDQ\cite{Goodman1997}), time-to-event outcomes in the absence of censoring (e.g., survival time), or duration of events (e.g., breastfeeding duration).  

When faced with this challenge in practice, the typical approach is to either ``ignore or transform" - ignore the skewness in the data entirely and continue to define the estimand as a contrast of expected potential outcomes, or transform the outcome to obtain a more symmetric distribution for which the expectation is interpretable as the centre value. Both approaches have their advantages, although neither is entirely satisfactory in many practical settings. Ignoring the skewness may be appropriate when the expected potential outcomes are of direct interest, allowing established confounding-methods to be applied\cite{Hernan2020}, although this may not reflect the central value of the outcome distribution. Transformation of the outcome to be more symmetrically distributed could be an alternative solution. However, this relies on a suitable transformation existing (which may not be feasible for highly skewed distributions) and makes interpretation of the causal effects more complex than interpretation in the original scale (e.g., log-years instead of years). 

An appealing alternative could be to define the causal effect using a contrast of median potential outcomes (i.e., the causal difference in medians). In fact, the causal effect has been generally defined as a contrast of any functional of the distributions of counterfactual outcomes under different exposure values.\cite{Hernan2004} However, despite being a widely acknowledged concept, there is limited availability and awareness of confounding-adjustment methods to estimate the causal difference in medians in practice. A handful of previous studies have acknowledged the need for such methods, and presented derivations of approaches to estimate causal effects as contrasts of distribution quantiles more generally. A study by Zhang et al. (2012) derived a number of methods - a quantile regression estimator, an inverse-probability weighted estimator, and a stratified estimator using propensity scores.\cite{Zhang2012} A more recent study in the field of environmental science defined a novel ``overlap weighting" estimator using a class of balancing weights from functions of the propensity score model to weight each group to a selected target population.\cite{Sun2021} An immediate suggestion may be to use multivariable quantile regression as explored previously\cite{Zhang2012, Sun2021}, although this relies on the strict constant-effect assumption which may be too simplistic in practice.

Despite these advances, application of these methods in epidemiological research remains scarce. The current discussion and evaluation of such methods is relatively limited, which has potentially led to a lack of awareness in their existence. Furthermore, to the best of our knowledge, the use of g-computation in the context of medians has not been widely discussed, let alone studied in relation to other approaches. In addition, there has been limited investigation of how these methods perform and compare in realistic settings across various scenarios, specifically in terms of the degree of skewness in the outcome which has only been explored minimally and not for all the methods considered here.\cite{Sun2021}

In this paper we aim to describe and compare confounding-adjustment methods to estimate the causal difference in medians, intending to increase understanding of their utility and encourage application in practice where appropriate. We begin this paper by defining the causal effect of interest alongside an illustrative example from the Longitudinal Study of Australian Children (LSAC)\cite{Sanson2002}, before outlining the confounding-adjustment methods considered. We then report findings from a simulation study motivated by the LSAC example, in addition to demonstration of the methods applied to the LSAC data. We conclude by summarising the key findings, strengths, limitations and practical recommendations. 


\section*{Defining the causal effect using medians}\label{sec:causal_effect}

Consider an observational study with continuous skewed outcome variable $Y$, a binary exposure variable $A$, and a vector of $K$ confounder variables $\bm{C}$. We assume $\bm{C}$ includes only binary or continuous variables, noting that categorical confounders can be represented as a set of binary indicators. For simplicity of discussion, we have restricted $A$ to be binary and assume that no variable is subject to missingness. 

The example used throughout this paper involves data from 4882 children from a longitudinal cohort study (LSAC).\cite{Sanson2002, Christensen2017} Children aged 4-5 years were recruited in 2004 (wave 1; approved by the Australian Institute of Family Studies Ethics Committee), with follow-ups every two years in subsequent waves. The example investigation examined the impact of maternal mental health on a child's behaviour in early childhood. The exposure ($A$) was a binary indicator of probable serious maternal mental illness, with the outcome ($Y$) being the child's behavioural difficulties as measured by the Strengths and Difficulties Questionnaire\cite{Goodman1997} (SDQ). Higher SDQ scores indicate increased behavioural difficulties, with scores being positively skewed in the LSAC cohort (see supplementary Figure S1). Potential confounders ($\bm{C}$) included demographic information about the child and mother (see Table \ref{tab:lsac_variables} for a full description of all variables). 

We define $Y^{a}$ to be the potential outcome when the exposure is set to level $a$. In the LSAC example, $Y^a$ represents the SDQ score for a child when their mother is set to have a probable serious mental illness ($a=1$) versus not ($a=0$). Here we define the shorthand notation $m_a$ to denote the median (\textit{med}) potential outcome under exposure level $A=a$, such that $m_a = med[Y^a]$ and $m_a \in \mathbb{R}$. Therefore, the causal difference in medians, denoted by $\delta$, is defined as the difference between the median potential outcomes under the two exposure levels:
\begin{eqnarray}
\delta = med[Y^{a=1}] - med[Y^{a=0}] = m_1 - m_0. 
\end{eqnarray}
For the LSAC example, $\delta$ represents the difference in median SDQ scores if all children were exposed to maternal mental health problems compared to if none of them were exposed. 

\begin{center}
	\begin{table}[t!]%
		\caption{Overview of variables from the Longitudinal Study of Australian Children (LSAC) example \cite{Sanson2002, Christensen2017} examining the impact of maternal mental health on a child's behavioural difficulties in early childhood. The exposure and confounders are recorded at wave 1 (2004), with the outcome variable recorded at wave 3 (2008).  \label{tab:lsac_variables}}
		\small\centering
		\begingroup \fontsize{9.5}{15}\selectfont
		\begin{tabular}{l l l}
			\toprule
			\textbf{Role} & \textbf{Variable} & \textbf{Values and additional details} \\ 
			\toprule
			Outcome $Y$ & Behavioural difficulties score & Range 0-40; Strengths \& Difficulties Questionnaire\cite{Goodman1997} \\
			Exposure $A$ & Probable serious maternal mental illness & Yes:$A=1$/No:$A=0$; Yes defined as a K10 score $<$ 4 \cite{Kessler2002, Andrews2001} \\
			Confounders $\bm{C}$ & Sex of child & Male/Female \\
			& Whether the child has siblings & Yes/No \\
			& Child's physical functioning score & Range 0-100; Pediatric Quality of Life Inventory\cite{Varni1999} \\
			& Behavioural difficulties score (baseline) & Range 0-40; Strengths \& Difficulties Questionnaire\cite{Goodman1997} \\
			& Maternal age & Recorded in years \\
			& Maternal smoking status & Yes/No \\
			& Maternal risky alcohol consumption & Yes/No; Yes defined as $>$ 2 standard alcoholic drinks per day \\
			& Maternal completion of high school & Yes/No \\
			& Family financial hardship score & Range 0-6 \\
			& Consistent parenting score & Range 1-5 \\
			\bottomrule
		\end{tabular}
		\endgroup
	\end{table}
\end{center}

\vspace{-1.54cm}

The causal difference in medians is identifiable from observational data under the assumptions of consistency, conditional exchangeability given $\bm{C}$ and positivity (assumptions 1-3; see supplementary material for further detail), as has been shown elsewhere.\cite{Greenland1986,Sun2021} Whether these assumptions hold in practice is a matter of debate, however, for the remainder of this paper we assume these conditions do hold.


\section*{Confounding-adjustment methods}\label{sec:methods}
Under the aforementioned assumptions, the causal difference in medians $\delta$ can in principle be estimated from observable data using methods that adjust for potential confounding.  Here we introduce the confounding-adjustment methods investigated in our study, focusing on their implementation in practice. 


\subsection*{Multivariable quantile regression} \label{sec:quant_reg}
Multivariable regression is a commmon approach to estimate the average causal effect, adjusting for confounding through conditioning on the confounders. To estimate $\delta$, a natural adaptation is to use quantile regression (QR); a method for modelling the quantiles of the distribution of a random variable conditional on a set of covariates.\cite{Koenker1978} When applying this approach, a QR model is fitted for the outcome variable $Y$ conditional on both the exposure $A$ and confounder variables $\bm{C}$, with the $\tau^{th}$ quantile of $Y$ modelled as
\begin{eqnarray}
Q_\tau(Y|A,\bm{C}) = \beta_0(\tau) + \beta_1(\tau)A + \bm{\beta_3}^\mathrm{T}(\tau)\bm{C},
\end{eqnarray}
By setting $\tau=0.5$, the coefficient of the exposure variable $\beta_1(0.5)$ encodes the difference in the conditional median outcome between exposure groups for every level of $\bm{C}$ (confounder strata), i.e., $med[Y|A=1, \bm{C}=\bm{c}] - med[Y|A=0, \bm{C}=\bm{c}]$. Under assumptions (1-3), the assumption of a constant causal effect across confounder strata and assuming the QR model is correctly specified, the estimated exposure coefficient $\hat{\beta_1}(0.5)$ has been shown to be a consistent estimator for the causal effect $\delta$.\cite{Sun2021} 

Quantile regression is a widely-applied and accessible method in practice, with implementation readily available in statistical software (e.g., the \textit{quantreg} package in R\cite{quantreg2019}). However, the assumption of a constant causal effect across confounder strata may be too simplistic. 


\subsection*{IPW estimator}
An alternative approach uses the framework of inverse probability weighting (IPW) to create a pseudo-population in which the distribution of $\bm{C}$ is balanced between exposure groups, such that the association between $A$ and $Y$ in the pseudo-population provides an unbiased estimate for the causal effect of $A$ on $Y$.\cite{Horvitz1952, Hernan2000} To create the pseudo-population, observations are re-weighted in a way that is inversely proportional to the probability of the observed exposure conditional on the confounding variables via calculation of inverse probability (IP) weights. These probabilities are estimated from a model for the propensity score (PS) defined as $\pi(\bm{c}) = P(A=1|\bm{C}=\bm{c})$.  

A study by Zhang et al. (2012) derived an IPW estimator for the causal difference in medians and other quantiles of the potential outcome distribution.\cite{Zhang2012} Applying this approach in the context of medians, $m_a$ is estimated as the solution to the equation
\begin{eqnarray}
\sum_{i=1}^n \hat{W}_{a,i} I(Y_i \leq m_a) = 0.5,
\label{eq:ipw_estimator}
\end{eqnarray}
where $\hat{W}_{a,i}$ denotes the estimated weight for observation $i=1,\ldots,n$ under exposure level $A_i=a$. The weights are defined as $W_{a,i}=I(A_i=a)/[nP(A_i=a|\bm{c}_i)]$ and calculated using estimates of the propensity score, $\hat{\pi}(\bm{c})$, obtained via a suitable regression model (e.g., a logistic regression model for $A$ conditional on $\bm{C}$). Following the suggestion of Zhang et al. (2012), normalised weights $\hat{W}^*_{a,i}$ are preferred to improve finite-sample performance\cite{Zhang2012}, and are calculated by dividing each weight by the sum of all weights in the associated exposure group (see supplementary material for mathematical formulation).

To obtain the estimates $\hat{m}_1$ and $\hat{m}_0$, Equation \ref{eq:ipw_estimator} is solved for each exposure level $a$. In practice, statistical software can be used to solve this equation (e.g., via the \textit{uniroot} function in R). Under assumptions (1-3) and assuming that the propensity score model is correctly specified, the difference between these two values consistently estimates the causal difference in medians $\delta$. 


\subsection*{Weighted quantile regression}
An alternative implementation of IPW uses IP weights to fit a weighted quantile regression (QR) model, weighting the score equations of the regression as opposed to the observed outcomes (as is done via the IPW estimator). Using this approach, a univariable QR model for the $\tau^{th}$ quantile of $Y$ is specified as
\begin{eqnarray}
Q_\tau(Y|A) = \beta_0^*(\tau) + \beta_1^*(\tau)A,
\end{eqnarray}
and fit using the estimated IP weights as outlined for the IPW estimator above. By setting $\tau=0.5$, the coefficient of the exposure variable encodes the difference in medians between each exposure level in the pseudo-population. Under assumptions (1-3) and assuming the propensity score model is correctly specified, the estimate $\hat{\beta_1^*}(0.5)$ has been shown to be a consistent estimator for the causal difference in medians $\delta$.\cite{Sun2021}

Unlike the previous IPW estimator, which needs to be hand-coded at present, implementation of weighted QR is readily available within software packages (e.g., the \textit{weights} argument within the \textit{rq} function in R\cite{quantreg2019, R2019}).


\subsection*{G-computation}

G-computation is another popular confounding-adjustment method, arising from the g-formula which states that under assumptions (1-3), the marginal density of $Y^a$ can be identified from observable data as\cite{Robins1986}
\begin{eqnarray} \label{eq:integral}
f_{Y^a}(y) = \sum_{\bm{c}} f_{Y|A,\bm{C}} (y|a,\bm{c}) f_{\bm{C}}(\bm{c}), 
\end{eqnarray}
which is equivalent to $\mathbb{E}[f_{Y|A,\bm{C}}(y|a,\bm{C})]$ where the outer expectation is over $\bm{C}$, where $f$ is defined as the density. Intuitively, the right-hand side is standardising the conditional density under an exposure value to the distribution of the confounders in the whole sample, thus addressing the imbalance in confounders between exposure groups due to non-randomisation, which enables a contrast under exposure values comparable. This result can be used to identify any functional of the marginal density of $Y^a$.   

When interest is in the median potential outcome $m_a$, the g-formula implies that $m_a$ can be identified as the solution to
\begin{equation}
\int_{-\infty}^{m_a} \mathbb{E}[f_{Y|A,\bm{C}}(y|a,\bm{C})] dy = 0.5. \label{eq:pdf2}
\end{equation}
Here we note the term inside the integral is the expectation of a density, and thus the median potential outcome is not identified by a simple aggregation across the sample (i.e., the median of the conditional medians or expectations). Instead, estimation of $m_a$ using g-computation requires estimation of the conditional density within the integral. Next we describe two possible implementations of g-computation for estimation of $m_a$ - a Monte Carlo integration-based approach and an approximate approach, denoted as \textit{g-comp (MC)} and \textit{g-comp (approx)}, respectively.

\subsubsection*{G-comp (MC)}
The first implementation, g-comp (MC), uses Monte Carlo simulation to perform draws from the density $f_{Y^a}(y)$, which can then be used to estimate $m_a$.\cite{Tsiatis2020} Specifically, we posit a model for the conditional density of $Y$ given $A=a$ and $\bm{C}$. We then repeatedly draw from the expectation over $\bm{C}$ of this density, i.e., corresponding to draws from $f_{Y^a}(y)$ (based on Equation \ref{eq:integral}). The median potential outcome under exposure value $a$ is estimated as the median of these draws.\cite{Tsiatis2020}  

Implementing this approach requires a model for the distribution of $Y$ conditional on $A$ and $\bm{C}$ (referred to as the outcome model). Given the skewed nature of the outcome in this setting, one possible approach is to assume that $Y \sim A,\bm{C}$ follows an approximate log-normal distribution, with the mean of the underlying normal distribution dependent on $A$ and $\bm{C}$. The g-comp (MC) method would therefore be implemented as follows:
\begin{enumerate}
	\item Fit a linear model for $\mathrm{log}(Y)$ conditional on $A$ and $\bm{C}$ to the observed data. 
	\item Using the fitted linear model, obtain predictions of the mean outcome (on the log scale) for every observation (where $i=1,...,n$) twice:
	\begin{enumerate}
		\item Set each observation to be exposed (i.e., set $A_i=1$) to obtain predictions $\hat{\mu}^1_i = \hat{\mathbb{E}}[\mathrm{log}(Y_i)|A_i=1, \bm{C}_i=\bm{c}_i]$, $i=1,...,n$.
		\item Set each observation to be unexposed (i.e., set $A_i=0$) to obtain predictions $\hat{\mu}^0_i = \hat{\mathbb{E}}[\mathrm{log}(Y_i)|A_i=0, \bm{C}_i=\bm{c}_i]$, $i=1,...,n$.
	\end{enumerate}
	\item For $a=1,0$, repeatedly perform $R$ draws for each observation from a log-normal distribution parametrised with the mean (on the log scale) equal to $\hat{\mu}^a_i$ and standard deviation $\hat{\sigma}$ equal to the estimated residual deviance of the model fitted in step 1.
	\item For $a=0,1$, the median of the combined $R$ samples drawn for each of the $n$ observations are used to obtain an estimate of the median potential outcome ($\hat{m}_a$).\cite{Tsiatis2020}
\end{enumerate}

\subsubsection*{G-comp (approx)}
In practice, the above approach may be computationally intensive, particularly for a data set with a large number of observations. Therefore, an alternative approach, g-comp (approx), approximates $\mathbb{E}[f_{Y|A,\bm{C}}(y|a,\bm{c})]$ by obtaining estimates of it across a grid of candidate $y$ values, denoted as $y^*$, and then solves Equation \ref{eq:pdf2} numerically to obtain the estimated median potential outcome. This approach again requires a model for the density and an assumed distribution for $Y \sim A, \bm{C}$ (e.g., log-normality). 

When implementing this approach, steps 1 and 2 are performed as outlined above. For $a=0,1$ and for each candidate $y^*$, we then estimate $\hat{f}_{Y|A,\bm{C}}(y^*|a, \bm{c}_i)$ for each record $i$ by assuming a log-normal density with the mean (on the log scale) equal to $\hat{\mu}^a_i$ and standard deviation $\hat{\sigma}$ (as defined in step 3 above). Averaging the conditional densities over the sample yields the estimated expectation for candidate value $y^*$:
\begin{eqnarray}
\hat{\mathbb{E}}[f_{Y|A,\bm{C}}(y^*|a,\bm{c})] = \frac{1}{n} \sum_{i=1}^n \hat{f}_{Y|A,\bm{C}}(y^*|a, \bm{c}_i).
\end{eqnarray}
By repeating this process for every candidate value $y^*$, the expectation within the integral in Equation \ref{eq:pdf2} is estimated across the range of $Y$. This integral is then approximated by adding up these values cumulatively, and finding the minimum value of $y^*$ for which this sum is equal to 0.5 to estimate $m_a$.  

For both g-computation implementations, the difference between the estimates $\hat{m_1}$ and $\hat{m_0}$ consistently estimates the causal difference in medians under assumptions (1-3) and assuming that the outcome model is correctly specified.\cite{Tsiatis2020} 

\subsection*{Standard error estimation}
For all confounding-adjustment methods outlined here, it is advised that standard errors and confidence intervals (CIs) are estimated using bootstrap procedures (using the percentile method) based on previous recommendations\cite{Zhang2012}.


\section*{Simulation Study}\label{sec:simulation_study}


A simulation study was conducted motivated by the LSAC example to investigate the performance of each confounding-adjustment method in a realistic setting and under varying degrees of skewness in the outcome variable.  


\subsection*{Design of the simulation study}\label{sec:results}

We generated 1000 datasets consisting of 1000 records for each of four skewness scenarios considered. For each scenario, dataset and record, five confounder variables $C_k$ for $k=1,\dots,5$ (three binary and two continuous), a binary exposure $A$ and a skewed continuous outcome $Y$ were generated based on variables in the LSAC data set (see supplementary Table S1 for a full outline of each variable and their generating distribution). Values for log($Y$) were generated from a normal distribution with the mean defined by a linear regression model including $A$ and $\bm{C}$ as predictors. Different skewed distributions in $Y$ were established by setting the standard deviation in the generating normal distribution for log($Y$) to $\sigma=0.75, 1, 1.25, 1.5$ signifying increasing skewness scenarios. Values for log($Y$) were exponentiated to obtain the outcome value $Y$, with the distribution of $Y$ being positively skewed (supplementary Figure S2). Here we note that $\sigma$ characterises other properties of the outcome distribution beyond the measure of skewness. Therefore, findings across skewness scenarios are not expected to behave in a monotonic pattern. 

Data was generated under two different confounding settings - weak confounding bias (approximately 10\% relative bias in the unadjusted estimate relative to the true value) and strong confounding bias (approximately 20\% relative bias). Modifications to coefficients in the outcome-generating model were used to achieve this.

The true causal difference in medians $\delta$ in each scenario was computed by empirical methods\cite{Sun2021} (see supplementary material for further details and true values used). Coefficients in the outcome-generating model were modified to ensure the true value was large enough to be estimable with the given sample size with adequate power (approximately 80\%) in an unadjusted analysis.

The confounding-adjustment methods were applied to each simulated dataset to estimate $\delta$, alongside an unadjusted contrast of sample medians across exposure groups. Specific details about model specifications and implementation are provided in supplementary material, although here we note that both the propensity score model and outcome model were correctly specified (i.e., consistent with the data generation approach). 

For each confounding strength and skewness scenario, metrics assessing the performance of each method were calculated using the formulae in Morris et al. (2019)\cite{Morris2019} and averaged over the 1000 samples, with key focus on bias in the point estimate. Monte Carlo standard errors were estimated for each metric. All analysis was conducted in R 4.0.2\cite{R2019}, using the \textit{quantreg}\cite{quantreg2019} and \textit{boot}\cite{boot1997,boot2022} packages within self-developed code (available at \url{https://github.com/daisyshep/CI-medians.git}), with the IPW estimator implemented using R code supplied in the method's source paper.\cite{Zhang2012}


\subsection*{Results from the simulation study}\label{sec:results}

\begin{figure}[t!]
	\centering
	\includegraphics[scale=0.58]{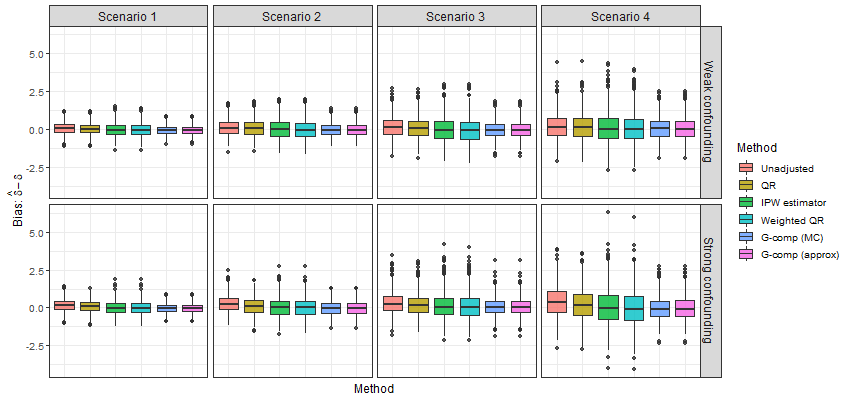}
	\caption{Calculated bias in causal difference estimates obtained under each method, skewness scenario (strength of skewness increases with scenario number, with scenario 1 and 4 corresponding to the weakest and highest skewness, respectively) and confounding setting in the simulated datasets (1000 datasets per skewness scenario) analysed under each method. \label{fig:sim_results}}
\end{figure}

Estimates for $\delta$ differed depending on the method used, with the variation in estimates tending to increase with increasing skewness for both confounding strengths (Figure \ref{fig:sim_results}). Estimates obtained using multivariable QR were biased across all skewness scenarios (relative bias range: 6.6\% to 8.1\% weak confounding, 9.0\% to 14.4\% strong confounding; Table \ref{tab:sim_results}); an expected result given the method's strict assumption of a constant causal effect across stratum which did not hold in the data generating mechanism. In contrast, methods which relaxed this assumption (the IPW estimator, weighted QR and both implementations of g-computation) performed well across both confounding strength settings, with minimal bias in estimates for $\delta$ (relative bias $<$ 5\% in the majority of skewness scenarios).

Both g-comp (MC) and g-comp (approx) had similar relative bias to one another across all simulation scenarios (i.e., differing by $<$0.1\% between implementations), with g-comp (approx) being quicker computationally to implement. Furthermore, both approaches yielded lower empirical standard errors comparatively to all other methods across all settings, therefore minimising both the bias and variance simultaneously.  

There was little bias in estimating the standard error (SE) with the bootstrap for all methods, with a slight overestimation for the IPW and weighted QR estimator. In all settings, the coverage probability was close to nominal level for all confounding-adjustment methods (range: 93.60\% to 97.70\%).

\section*{Application to LSAC}\label{sec:lsac}

\begin{figure}[t!]
	\centering
	\includegraphics[scale=0.65]{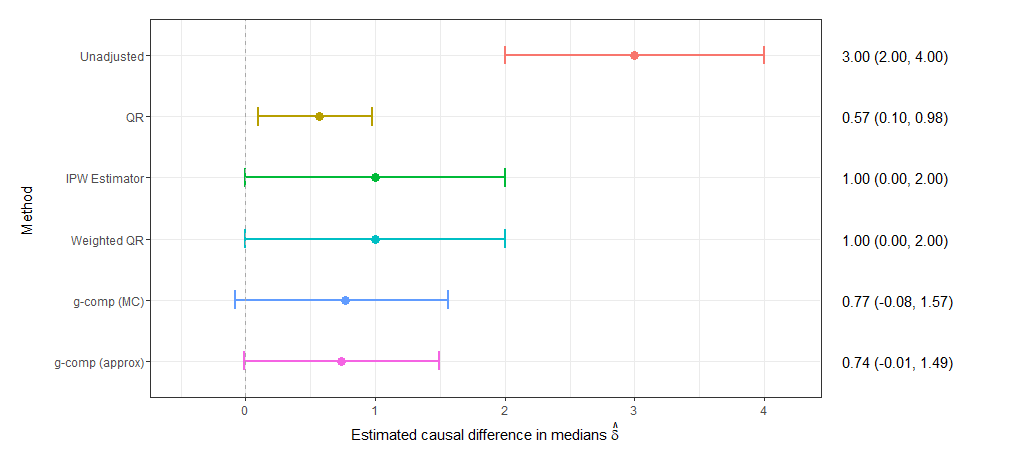}
	\caption{Estimated causal difference in medians $\hat{\delta}$ for the Longitudinal Study Of Australian Children (LSAC) example\cite{Sanson2002, Christensen2017} obtained under each confounding-adjustment method, where $\hat{\delta} = \hat{m}_1 - \hat{m}_0$ is the estimated difference in median SDQ scores if all children were exposed to maternal mental health problems compared to if none of them were exposed. Point estimates and their corresponding 95\% confidence intervals are presented alongside the figure.\label{fig:lsac_results}}
\end{figure}

Methods were applied to the LSAC data using complete cases only ($n=3245$), with specific details about their implementation provided in supplementary material. All methods estimated the median SDQ score for children exposed to maternal mental health problems to be higher than the median SDQ score for children if they were not exposed (Figure \ref{fig:lsac_results}), suggesting moderately increased behavioural problems in early childhood for children of mothers with mental illness. Estimated effects were more consistent across the methods than observed within the simulation study, although multivariable QR yielded lower estimates of $\delta$ than the other methods. The IPW estimator and weighted QR produced the same point estimates and bootstrap CIs to one another, as did the g-computation approaches, which estimated a slightly smaller causal effect comparatively.

\begin{table}[htbp]
	\caption{Performance of confounding-adjustment methods across confounding and skewness scenarios in the simulation study with maximum Monte Carlo standard errors (SE) provided in the table footnote (see supplementary Table S3 for all Monte Carlo SEs). \label{tab:sim_results}}
	\centering
	\begingroup \fontsize{9}{12}\selectfont
	\begin{tabular}{p{1.9cm} p{1.4cm} r >{\raggedleft\arraybackslash}p{1.2cm} >{\raggedleft\arraybackslash}p{1.5cm} >{\raggedleft\arraybackslash}p{1.5cm} >{\raggedleft\arraybackslash}p{1.25cm} >{\raggedleft\arraybackslash}p{1.5cm} >{\raggedleft\arraybackslash}p{1.5cm}}
		\toprule
		\textbf{Confounding} & \textbf{Skewness scenario} & \textbf{Method} & \textbf{Bias} & \textbf{Relative bias} (\%) & \textbf{Empirical SE} & \textbf{Model SE} & \textbf{Error SE} (\%) & \textbf{Coverage} (\%) \\ 
		\toprule
		
		Weak & 1 & Unadjusted & 0.090 & 10.07 & 0.369 & 0.381 & 3.21 & 94.60 \\ 
		& & QR & 0.059 & 6.63 & 0.384 & 0.375 & -2.24 & 94.80 \\ 
		& & IPW estimator & -0.004 & -0.48 & 0.442 & 0.454 & 2.78 & 94.90 \\ 
		& & Weighted QR & -0.023 & -2.53 & 0.440 & 0.453 & 2.98 & 95.20 \\ 
		& & G-comp (MC) & -0.026 & -2.91 & 0.308 & 0.298 & -3.18 & 93.70 \\ 
		& & G-comp (approx) & -0.026 & -2.90 & 0.308 & 0.298 & -3.16 & 93.60 \\ \hline

		Weak & 2 & Unadjusted & 0.123 & 10.08 & 0.526 & 0.528 & 0.32 & 94.90 \\ 
		& & QR & 0.099 & 8.13 & 0.540 & 0.517 & -4.35 & 94.20 \\ 
		& & IPW estimator & 0.050 & 4.09 & 0.625 & 0.640 & 2.43 & 93.70 \\ 
		& & Weighted QR & 0.023 & 1.91 & 0.621 & 0.636 & 2.54 & 94.20 \\ 
		& & G-comp (MC) & 0.006 & 0.48 & 0.436 & 0.420 & -3.60 & 94.00 \\ 		
		& & G-comp (approx) & 0.006 & 0.45 & 0.436 & 0.420 & -3.62 & 93.90 \\ \hline

		Weak & 3 & Unadjusted & 0.160 & 10.00 & 0.684 & 0.727 & 6.27 & 95.60 \\ 
		& & QR & 0.114 & 7.14 & 0.693 & 0.695 & 0.31 & 94.70 \\ 
		& & IPW estimator & 0.019 & 1.17 & 0.806 & 0.870 & 7.94 & 95.00 \\ 
		& & Weighted QR & -0.020 & -1.23 & 0.803 & 0.864 & 7.62 & 95.20 \\ 
		& & G-comp (MC) & -0.016 & -1.00 & 0.550 & 0.563 & 2.47 & 95.20 \\ 	
		& & G-comp (approx) & -0.015 & -0.97 & 0.550 & 0.563 & 2.40 & 95.10 \\ \hline

		Weak & 4 & Unadjusted & 0.193 & 10.09 & 0.833 & 0.894 & 7.24 & 96.80 \\ 
		& & QR & 0.133 & 6.95 & 0.864 & 0.842 & -2.62 & 95.50 \\ 
		& & IPW estimator & 0.123 & 6.45 & 1.032 & 1.098 & 6.40 & 95.30 \\ 
		& & Weighted QR & 0.081 & 4.22 & 1.017 & 1.091 & 7.24 & 95.50 \\ 
		& & G-comp (MC) & 0.062 & 3.25 & 0.695 & 0.712 & 2.40 & 95.00 \\  		
		& & G-comp (approx) & 0.063 & 3.28 & 0.694 & 0.710 & 2.20 & 94.90 \\ \hline

		Strong & 1 & Unadjusted & 0.171 & 20.17 & 0.391 & 0.398 & 1.66 & 93.20 \\ 
		& & QR & 0.098 & 11.58 & 0.398 & 0.390 & -2.10 & 94.30 \\ 
		& & IPW estimator & 0.003 & 0.31 & 0.441 & 0.466 & 5.75 & 95.40 \\ 
		& & Weighted QR & -0.017 & -1.95 & 0.440 & 0.464 & 5.33 & 95.30 \\ 
		& & G-comp (MC) & -0.023 & -2.65 & 0.308 & 0.306 & -0.93 & 94.10 \\ 	
		& & G-comp (approx) & -0.023 & -2.67 & 0.308 & 0.306 & -0.86 & 94.10 \\ \hline

		Strong & 2 & Unadjusted & 0.240 & 20.05 & 0.544 & 0.568 & 4.40 & 94.00 \\ 
		& & QR & 0.117 & 9.81 & 0.556 & 0.556 & -0.03 & 95.00 \\ 
		& & IPW estimator & 0.030 & 2.55 & 0.621 & 0.673 & 8.34 & 96.00 \\ 
		& & Weighted QR & 0.004 & 0.33 & 0.617 & 0.670 & 8.61 & 95.40 \\ 
		& & G-comp (MC) & -0.025 & -2.08 & 0.446 & 0.441 & -1.22 & 94.30 \\ 	
		& & G-comp (approx) & -0.025 & -2.10 & 0.446 & 0.440 & -1.22 & 94.00 \\ \hline

		Strong & 3 & Unadjusted & 0.307 & 20.11 & 0.729 & 0.734 & 0.72 & 93.50 \\ 
		& & QR & 0.219 & 14.39 & 0.726 & 0.709 & -2.32 & 93.70 \\ 
		& & IPW estimator & 0.117 & 7.68 & 0.847 & 0.897 & 5.89 & 95.40 \\ 
		& & Weighted QR & 0.082 & 5.35 & 0.843 & 0.891 & 5.77 & 95.50 \\ 
		& & G-comp (MC) & 0.076 & 4.97 & 0.570 & 0.579 & 1.62 & 95.10 \\ 	
		& & G-comp (approx) & 0.075 & 4.93 & 0.569 & 0.579 & 1.75 & 95.20 \\ \hline

		Strong & 4 & Unadjusted & 0.420 & 20.02 & 1.004 & 1.052 & 4.76 & 95.10 \\ 
		& & QR & 0.189 & 8.98 & 1.002 & 0.993 & -0.88 & 94.50 \\ 
		& & IPW estimator & 0.038 & 1.81 & 1.189 & 1.252 & 5.37 & 95.50 \\ 
		& & Weighted QR & -0.009 & -0.43 & 1.180 & 1.241 & 5.19 & 95.50 \\ 
		& & G-comp (MC) & -0.028 & -1.34 & 0.759 & 0.806 & 6.11 & 95.40 \\ 		
		& & G-comp (approx) & -0.028 & -1.32 & 0.760 & 0.791 & 4.12 & 95.70 \\ 
		\bottomrule
	\end{tabular}
	\begin{tablenotes}
		\item Maximum Monte Carlo SE (performance measure): 0.038 (bias), 0.018 (relative bias), 0.010 (empirical SE), 
		0.029 (model SE), 6.956\% (relative error in model SE), 0.796\% (coverage). 
	\end{tablenotes}
	\endgroup
\end{table}


\section*{Discussion}\label{sec:conclusion}

In the presence of skewed outcome data, the common approach to ``ignore or transform" may not be optimal, and defining the causal effect using a contrast of median potential outcomes may be more appropriate. Despite being a widely acknowledged concept, there is scarce availability and awareness of confounding-adjustment methods to estimate this parameter in practice. A handful of previous studies have proposed approaches to estimate causal effects as contrasts of distribution quantiles more generally\cite{Sun2021, Zhang2012}, but investigation of these methods and their application in health and medical studies remain scarce. 

In this paper we aimed to address this gap by describing and evaluating methods identified from previous literature (multivariable quantile regression\cite{Sun2021}, an IPW estimator\cite{Zhang2012} and weighted quantile regression\cite{Sun2021}) alongside two implementations of g-computation that, to the best of our knowledge, have not been widely described or studied alongside other methods. The confounding-adjustment methods were selected and described with a key focus on their accessibility and ease of implementation in a practical setting, with code made available, to encourage their use in practice where applicable. 

Results from the simulation study indicated varied performance of the confounding-adjustment methods when estimating the causal difference in medians. As anticipated, the multivariable QR was too simplistic for the realistic setting reflected in our simulated datasets and produced biased estimates. The IPW estimator, weighted QR and both implementations of g-computation yielded estimates with minimal bias, with g-computation additionally minimising the variance in estimates; an expected observation as IPW estimates tend to be more variable than those obtained via g-computation.\cite{Moreno2017} We also note that the g-comp (approx) implementation was computationally more efficient than g-comp (MC) provided a suitable $y^*$ range is used. 

These findings need to be interpreted in light of the fact that under our data generation approach, both the propensity score model (used for the IPW estimator and weighted QR) and outcome model (used for g-computation) were correctly specified; a critical assumption when applying the singly robust methods as noted by Zhang et al. (2012) in the case of IPW.\cite{Zhang2012} However, in the context of the medians, the outcome model for g-computation relates to specifying a model for the whole density. In practice, correctly specifying this model may be harder to achieve than a correctly specified propensity score model, and thus could be considered a stronger assumption than for the weighted methods.   

A strength of this work was the design of our simulation study motivated by the LSAC example, allowing us to investigate the performance of these methods in a realistic scenario. Further we investigated varying skewness distributions in the outcome variable, alongside two different strengths of confounding bias present in the datasets, resulting in a more complex and realistic study than those explored in previous papers.\cite{Zhang2012} Additionally, our inclusion of the g-computation approach (under two implementations) in an accessible and clear manner, has brought light to a little discussed approach.  

A potential limitation of our study was the restriction to singly robust methods only, which rely on correct specification of the respective model. Previous studies have presented a handful of promising doubly robust methods, which combine both a model for the outcome and a model for the exposure and rely on only one of the models being correctly specified to obtain an unbiased estimator.\cite{Zhang2012,Xie2020,Xu2018,Diaz2017} Doubly robust methods have not yet been evaluated in a range of complex and realistic scenarios as is done here for singly robust methods, with their implementation not as accessible or readily available in software and therefore their application remains scarce in practice. Both of these factors would be useful to pursue in future work. We also note the outcome distributions explored within this study were all heavily right-skewed, so results may represent an extreme representation of the methods' performance under heavily skewed outcome data.

In conclusion, when estimating the causal difference in medians the IPW estimator, weighted QR or g-computation present promising approaches, provided a richly specified model is used such that correct specification of the propensity score model or outcome model is likely. Implementation of the IPW estimator and weighted QR methods is readily available and therefore accessible to implement (e.g., \textit{rq} function in R\cite{R2019}, open-source code of the IPW estimator\cite{Zhang2012}). Implementations of the g-computation approach are not as readily available, but we have provided source code which can guide practitioners in the implementation of this method. Overall, these methods provide appealing alternatives to the common ``ignore or transform" approach or the stringent constant-effect assumption of multivariable quantile regression, enhancing our capability to obtain meaningful causal effect estimates with skewed outcome data. 


\subsection*{Author contributions}
Development of the project and study design was conducted by DS and MMB. DS conducted the simulation study and data analysis, with regular input from MMB. The manuscript was developed by DS, with critical input provided by MMB. BB provided input into the g-computation methodology, and feedback on the manuscript. 


\subsection*{Financial disclosure}
DS was supported by the Lorenzo and Pamela Galli Medical Research Trust. MMB was supported by an Australian Research Council Discovery Early Career Researcher Award (project number DE190101326) funded by the Australian Government and a NHMRC Investigator Grant (ID 2009572). 

\subsection*{Data availability statement}
Data used for the illustrative example within this paper are available from the Department of Social Services (DSS) with access provided by the National Centre for Longitudinal Data (NCLD). Restrictions apply to the availability of these data, which were used under license for this study. Data are available at \url{https://dataverse.ada.edu.au/dataset.xhtml?persistentId=doi:10.26193/F2YRL5} with the permission of the DSS.

\subsection*{Conflict of interest}
The authors declare no potential conflict of interests.


\bibliography{medians_bib}


\pagebreak

\baselineskip=20pt

\begin{center}
	{\bf \Large Confounding-adjustment methods for the causal difference in medians}
\end{center}

\begin{center}
	{\bf \large Supplementary Material}
\end{center}

\medskip

\begin{center}
	{\bf Daisy A. Shepherd*$^{1,2}$, Benjamin R. Baer$^{3}$ \& Margarita Moreno-Betancur$^{1,2}$}
\end{center}

\medskip

\noindent {\sl \rf \small $^{1}$Clinical Epidemiology \& Biostatistics Unit, Department of Paediatrics, The University of Melbourne, VIC, Australia \\
	$^{2}$Clinical Epidemiology \& Biostatistics Unit, The Murdoch Children's Research Institute, The Royal Children's Hospital, VIC, Australia \\
	$^{3}$Department of Biostatistics and Computational Biology, The University of Rochester, New York, USA}

\medskip 
\noindent {\bf*}Corresponding author: Daisy A. Shepherd. Email: daisy.shepherd@mcri.edu.au

\bigskip

\noindent {\rf \textbf{Abbreviations}: LSAC, The Longitudinal Study of Australian Children; SDQ, Strengths and Difficulties Questionnaire; QR, quantile regression; IP, inverse probability; IPW, inverse probability weighted; PS, propensity score}

\bigskip
\bigskip
\bigskip

\noindent {\large \textbf{Contents}} \medskip  \newline
\textbf{S1 \ \  The LSAC case study} \smallskip \newline
\textbf{S2 \ \ Defining the causal effect} \newline
\indent  S2.1 \ \ Identifiability assumptions \newline
\textbf{S3 \ \ IPW estimator} \newline
\indent  S3.1 \ \ Normalised weights \newline
\textbf{S4 \ \ Additional simulation study details} \newline
\indent  S4.1 \ \ Data generation \newline
\indent  S4.2 \ \ Distribution of the outcome variable \newline
\indent  S4.3 \ \ True causal difference in medians \newline
\indent  S4.4 \ \ Implementation of confounding-adjustment methods \newline
\indent S4.5 \ \ Monte Carlo standard errors \newline
\textbf{S5 \ \ Implementation of confounding-adjustment methods in LSAC example} \newline


\pagebreak
\setcounter{equation}{0}
\setcounter{figure}{0}
\setcounter{table}{0}
\makeatletter

\renewcommand*\thesection{S\arabic{section}}
\renewcommand*\thetable{S\arabic{table}}
\renewcommand*\thefigure{S\arabic{figure}}
\renewcommand*\theequation{S.\arabic{equation}}

\section*{S1 \ \ The LSAC case study}

\begin{figure}[htbp]
	\centering
	\includegraphics[scale=0.45]{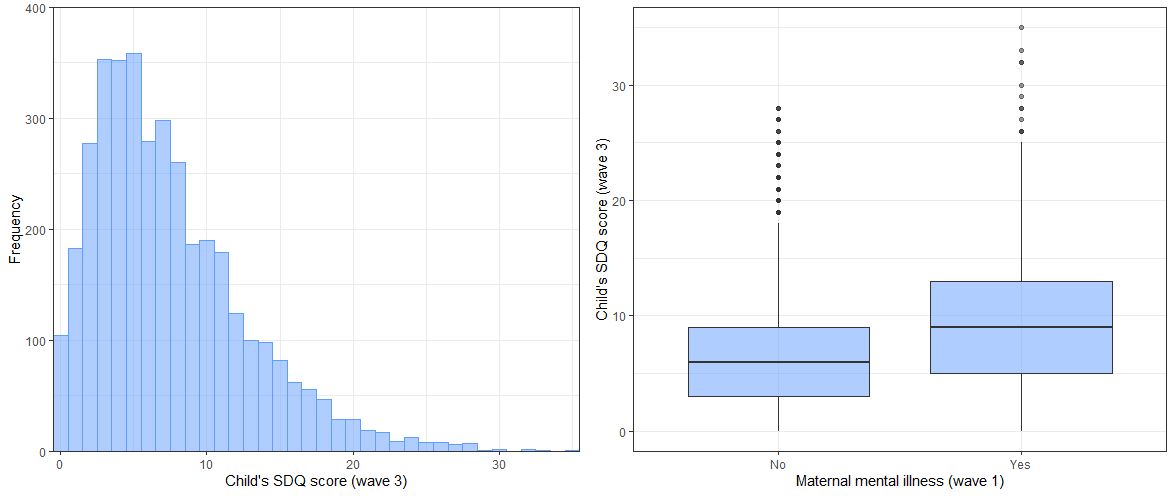}
	\caption{Distribution of the Strengths \& Difficulties questionnaire (SDQ) scores at wave 3 for the Longitudinal Study Of Australian Children (LSAC) case study.\cite{Sanson2002, Christensen2017}\label{fig:LSAC_outcome}}
\end{figure}

\section*{S2 \ \ Defining the causal effect}
To understand the formulation involved for the confounding-adjustment methods, we add further detail for the definition of the causal effect. Here we note that the median potential outcome under exposure level $A=a$ is defined as the solution to
\begin{eqnarray} \label{eq:pdf}
\int_{-\infty}^{m_a} f_{Y^a}(y) = 0.5,
\end{eqnarray}
where $f_{Y^a}(y)$ is the density function of $Y^{a}$, or alternatively as the $50^{\mathrm{th}}$ centile of the cumulative distribution function (CDF) of $Y^{a}$, $F_{Y^a}(y)=P(Y^a \leq y)$. Under the assumption that the CDF is continuous and strictly increasing for $a \in {0,1}$, then $m_a = F^{-1}_{Y^{a}}(0.5)$. Therefore the causal difference in medians, denoted by $\delta$, is defined as the difference between the median (denoted as $med$) potential outcomes under the two exposure levels:
\begin{eqnarray}
\delta = F^{-1}_{Y^{a=1}}(0.5) - F^{-1}_{Y^{a=0}}(0.5) = med[Y^{a=1}] - med[Y^{a=0}] = m_1 - m_0. 
\end{eqnarray}

\subsection*{S2.1 \ \ Identifiability assumptions \label{sec:ic}} 
The causal difference in medians is identifiable from observational data under the following identifiability assumptions. Firstly we require the consistency assumption, which states that the exposure $A$ corresponds to a well-defined intervention that in turn corresponds to the versions of the exposure in the data.\cite{Hernan2020} Under these conditions, the potential outcome $Y^a$ is well-defined and would be equal to $Y$ if an individual received exposure level $A=a$ (assumption 1). Secondly, we require the conditional exchangeability assumption given the selected set of confounders, which states that the potential outcome $Y^a$ is independent of the received exposure $A$ given $\bm{C}$, i.e., $Y^a \perp\!\!\!\perp A|\bm{C}$ (assumption 2). Thirdly, we require the positivity assumption, which states that every individual in the population has a positive probability of being exposed or unexposed, that is $P(A=a|\bm{C}=\bm{c}) > 0$, for all $\bm{c}$ with positive probability of occurring (assumption 3).

\bigskip

\section*{S3 \ \ IPW estimator}

Here we provide further explanation and details supporting the IPW estimator outlined in the main manuscript. The IPW estimator in Equation \ref{eq:ipw_estimator} from the main manuscript can be derived based on the following reasoning. Given assumptions (1-3) hold (as outlined in Section S2.1 above), then for a given exposure level $a$ the cumulative distribution function is equal to\cite{Zhang2012}
\begin{eqnarray}
F_{Y^a}(y) = \mathbb{E} \left[ \frac{I(A=a)I(Y \leq y)}{P(A=a|\bm{C})} \right]. 
\end{eqnarray}
Under each exposure level $a\in\{0,1\}$ and given estimates of the denominator probabilities from the fitted propensity score model $\pi{(\bm{c})}$, the expectation can be estimated using the sample average\cite{Horvitz1952, Robins1994} and can be regarded as the weighted empirical distribution of $Y$.\cite{Zhang2012} The IPW estimator of the median outcome value $m_a$ is therefore defined as the solution to Equation 3 in the main manuscript. It is important to note that Equation 3 may not have a unique solution, and therefore $m_a$ can be estimated by the value which minimises the difference between the two sides.\cite{Zhang2012}

\subsubsection{S3.1 \ \ Specification of normalised weights}
In place of weights $W_{a,i}$ in Equation 3, normalised weights ${W}^*_{a,i}$ are advised to be use to improve finite-sample performance.\cite{Zhang2012} These normalised weights are calculated by dividing each weight by the sum of all weights in the associated exposure group, such that
\begin{eqnarray}
{W}^*_{a,i} = \left. \frac{I(A_i=a)}{P(A_i=a|\bm{c}_i)} \ \right/ \sum_{i=1}^{n}\frac{I(A_i=a)}{P(A_i=a|\bm{c}_i)},
\end{eqnarray}
with the estimated values $\hat{W}^*_{a,i}$ replacing $\hat{W}_{a,i}$ in Equation 3 (in the main manuscript).


\section*{S4 \ \ Additional simulation study details}


\subsection*{S4.1 \ Data generation}

For each scenario, dataset and record, five confounder variables $C_k$ for $k=1,\dots,5$ (three binary and two continuous) were generated sequentially in order based on variables in the LSAC data set. Observations for the binary confounders $C_1, C_3$ and $C_4$ (based on sex, maternal education and financial hardship, respectively) were generated from a binomial distribution with the success probability defined by a logistic regression model using all previous confounders as predictors. For the continuous confounders $C_2$ and $C_5$ (maternal age and log-transformed baseline SDQ, respectively) values were generated from normal distributions with the means defined by a linear regression model including all previous confounders as predictors. Here we note none of these models included interaction terms, as outlined in Table \ref{tab:datagen_tab} below. 

\begin{figure}[htbp]
	\centering
	\includegraphics[scale=0.7]{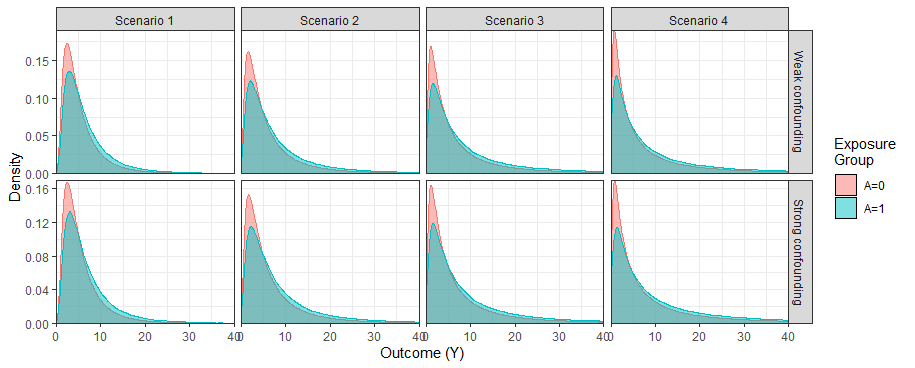}
	\caption{Distribution of the outcome variable ($Y$) under each skewness scenario and confounding bias strength used within the simulation study. \label{fig:skew_true}}
\end{figure}

Secondly, a binary exposure $A$ was generated for each record from a binomial distribution with success probability defined by a logistic regression model including all confounders as main effects. A skewed continuous outcome $Y$ was then generated for each record. Values for log($Y$) were generated from a normal distribution with the mean defined by a linear regression model including the exposure and all confounders as predictors. The linear regression model was specified to include all main effects and two exposure-confounder interaction terms based on interactions observed in the LSAC example. Different skewed distributions in the outcome variable were established by setting the standard deviation in the generating normal distribution for log($Y$) to $\sigma=0.75, 1, 1.25, 1.5$ for each of the four increasing skewness scenarios, respectively. Values for log($Y$) were then exponentiated to obtain the outcome value $Y$, with the distribution of $Y$ being positively skewed (Figure \ref{fig:skew_true}).

\begin{table}[htbp]
	\caption{Detail on the models used to generate variables for the simulation study, with the structure based on the LSAC dataset unless otherwise specified.  \label{tab:datagen_tab}}
	\begingroup\fontsize{9}{14}\selectfont
	\begin{tabular}{p{1.5cm} l l p{7cm}}
		\hline
		\textbf{Variable} & \textbf{LSAC variable} & \textbf{Generating distribution} & \textbf{Additional details} \\
		\hline
		$C_1$ & Sex & $C_1 \sim$ Binomial(1, 0.51) & \\
		$C_2$ & Maternal age & $C_2 \sim$ Normal(35.17, 5.47) & \\
		$C_3$ & Maternal education & $C_3 \sim$ Binomial(1, $Pr(C_3=1)$) & logit$[Pr(C_3=1)] = -1.41 + 0.78C_1 + 0.04C_2$ \\
		$C_4$ & Financial hardship$^a$ & $C_4 \sim$ Binomial(1, $Pr(C_4=1)$) & logit$[Pr(C_4=1)] = -1.55 + 0.47C_1 + 0.03C_2 + 0.80C_3$ \\
		$C_5$ & Baseline SDQ (logged) & $C_5 \sim$ Normal($\mu_{C_5}$, 0.63) & $\mu_{C_5} = 1.91 + 0.03C_1 + 0.01C_2 + 0.05C_3 + 0.12C_4$ \\
		$A$ & Maternal mental health & $A \sim$ Binomial(1, $Pr(A=1)$) & logit$[Pr(A=1)] = -2.39 + 0.04C_1 - 0.05C_2 - 0.09C_3 + 0.51C_4 + 1.07C_5$\\
		log($Y$) & SDQ score & log($Y$) $\sim$ Normal($\mu_Y, \sigma$) & $\mu_Y = 1.40 + 0.49A + 0.03C_1 - 0.01C_2 + 0.01C_3 + 0.03C_4 + 0.26C_5 + 0.12AC_1 - 0.01AC_2$ \\
		& & & $\sigma=0.75, 1, 1.25, 1.5$\\
		\hline
	\end{tabular} \\
	$^a$Dichotomised: No (score = 0 on original scale), Yes (score $>$ 0 on original scale)
	\endgroup
	
\end{table}

\newpage


\subsection*{S4.2 \ Distribution of the outcome variable}

\subsection*{S4.3 \ True causal difference in medians} \label{app:true}

Within the simulation study, the true causal difference in medians $\delta$, with respect to which assess bias, was computed by empirical methods outlined in the supplementary material of Sun et al. (2021).\cite{Sun2021} Specific details of our implementation are as follows. Initially we generated a large dataset (1,000,000 observations) per skewness scenario and strength of confounding bias (weak or strong). For each large dataset, a quantile regression model (including all main effects and exposure-confounder interactions) was fitted under different quantiles $\tau$; we used 200 different values for $\tau$ equally distributed across the range [0.05, 0.95]. For our vector of $m_a$ support values, we covered the range [3,8] in increments of 0.005. Different values of $\delta$ were obtained for each skewness scenario and under the two different strengths of confounding bias, as outlined below in Table \ref{tab:true_value}.\\

\begin{table}[htbp!]
	\centering
	\caption{True values for the causal difference in medians $\delta$ obtained under each skewness scenario and confounding setting. \label{tab:true_value}}
	\begin{tabular}{ccccc}
		\toprule
		\textbf{Confounding bias} & \multicolumn{4}{c}{\textbf{Skewness scenario}} \\ 
		\textbf{strength} & \textbf{1} & \textbf{2} & \textbf{3} & \textbf{4} \\
		\toprule
		Weak & 0.895 & 1.220 & 1.600 & 1.910 \\
		Strong & 0.850 & 1.195 & 1.525 & 2.100 \\
		\bottomrule
	\end{tabular}
\end{table}



\subsection*{S4.4 \ Implementation of confounding-adjustment methods in simulation study}

Here we outline the model specification and implementation of each confounding-adjustment method investigated in our study. For the multivariable quantile regression approach, the outcome model included $A$ and $\bm{C}$ as predictors with main effect terms only, as is the default in most software. For both the IPW estimator and the weighted quantile regression approaches, the propensity score model regressed $A$ on the confounders including main effect terms only. For the g-computation approaches, the outcome model was specified as $\mathrm{log}(Y)$ conditional on $A$ and $\bm{C}$, including main effect terms and two exposure-confounder interaction terms. Here we note that both the propensity score model and the outcome model were correctly specified (i.e., consistent with the data generation approach). For g-comp (MC), we performed $R=1000$ draws per observation. For g-comp (approx), the candidate $y^*$ values ranged over the values [0.01, 8] in increments of 0.01.


\newpage
\subsection*{S4.5 \ Monte Carlo standard errors}

\begin{table}[htbp!]
	\centering
	\caption{Monte Carlo standard errors of performance estimates calculated over the 1000 simulated datasets per skewness scenario and confounding bias strength under each of the confounding-adjustment methods. \label{tab:mc_results}}	
	\begingroup\fontsize{9}{11.3}\selectfont
	\begin{tabular}{p{2cm} p{1.35cm} l >{\raggedleft\arraybackslash}p{1cm} >{\raggedleft\arraybackslash}p{1.4cm} >{\raggedleft\arraybackslash}p{1.5cm} >{\raggedleft\arraybackslash}p{1.5cm} >{\raggedleft\arraybackslash}p{1.5cm} >{\raggedleft\arraybackslash}p{1.4cm}}
		\toprule
		\textbf{Confounding} & \textbf{Skewness scenario} & \textbf{Method} & \textbf{Bias} & \textbf{Relative Bias} & \textbf{Empirical SE} & \textbf{Model SE} & \textbf{Relative error SE (\%)} & \textbf{Coverage (\%)} \\ 
		\toprule
		Weak & 1 & Unadjusted & 0.012 & 0.013 & 0.008 & 0.002 & 2.387 & 0.715 \\ 
		&  & QR & 0.012 & 0.014 & 0.009 & 0.002 & 2.234 & 0.702 \\ 
		&  & IPW estimator & 0.014 & 0.016 & 0.010 & 0.003 & 2.414 & 0.696 \\ 
		&  & Weighted QR & 0.014 & 0.016 & 0.010 & 0.003 & 2.420 & 0.676 \\ 
		&  & G-comp (MC) & 0.010 & 0.011 & 0.007 & 0.001 & 2.179 & 0.768 \\ 
		&  & G-comp (approx) & 0.010 & 0.011 & 0.007 & 0.001 & 2.179 & 0.774 \\ \hline		
		
		Weak & 2 & Unadjusted & 0.017 & 0.014 & 0.008 & 0.004 & 2.607 & 0.715 \\ 
		&  & QR & 0.017 & 0.014 & 0.009 & 0.003 & 2.364 & 0.702 \\ 
		&  & IPW estimator & 0.020 & 0.016 & 0.010 & 0.007 & 2.768 & 0.696 \\ 
		&  & Weighted QR & 0.020 & 0.016 & 0.010 & 0.007 & 2.750 & 0.676 \\ 
		&  & G-comp (MC) & 0.014 & 0.011 & 0.007 & 0.002 & 2.234 & 0.768 \\ 
		&  & G-comp (approx) & 0.014 & 0.011 & 0.007 & 0.002 & 2.234 & 0.774 \\ \hline		
		
		Weak & 3 & Unadjusted & 0.022 & 0.014 & 0.008 & 0.009 & 3.352 & 0.715 \\ 
		&  & QR & 0.022 & 0.014 & 0.009 & 0.007 & 2.783 & 0.702 \\ 
		&  & IPW estimator & 0.025 & 0.016 & 0.010 & 0.013 & 3.735 & 0.696 \\ 
		&  & Weighted QR & 0.025 & 0.016 & 0.010 & 0.013 & 3.744 & 0.676 \\ 
		&  & G-comp (MC) & 0.017 & 0.011 & 0.007 & 0.003 & 2.426 & 0.768 \\
		&  & G-comp (approx) & 0.017 & 0.011 & 0.007 & 0.003 & 2.427 & 0.774 \\ \hline		
		
		Weak & 4 & Unadjusted & 0.026 & 0.014 & 0.008 & 0.014 & 4.412 & 0.715 \\ 
		&  & QR & 0.027 & 0.014 & 0.009 & 0.010 & 3.405 & 0.702 \\ 
		&  & IPW estimator & 0.033 & 0.017 & 0.010 & 0.022 & 5.563 & 0.696 \\ 
		&  & Weighted QR & 0.032 & 0.017 & 0.010 & 0.022 & 5.590 & 0.676 \\ 
		&  & G-comp (MC) & 0.022 & 0.012 & 0.007 & 0.006 & 2.947 & 0.768 \\ 
		&  & G-comp (approx) & 0.022 & 0.011 & 0.007 & 0.006 & 2.872 & 0.774 \\ \toprule
		
		Strong & 1 & Unadjusted & 0.012 & 0.015 & 0.009 & 0.002 & 2.357 & 0.796 \\ 
		&  & QR & 0.013 & 0.015 & 0.009 & 0.002 & 2.242 & 0.733 \\ 
		&  & IPW estimator & 0.014 & 0.016 & 0.010 & 0.003 & 2.491 & 0.662 \\ 
		&  & Weighted QR & 0.014 & 0.016 & 0.010 & 0.003 & 2.480 & 0.669 \\ 
		&  & G-comp (MC) & 0.010 & 0.011 & 0.007 & 0.001 & 2.228 & 0.745 \\ 
		&  & G-comp (approx) & 0.010 & 0.011 & 0.007 & 0.001 & 2.230 & 0.745 \\ \hline		
		
		Strong & 2 & Unadjusted & 0.017 & 0.014 & 0.009 & 0.005 & 2.582 & 0.796 \\ 
		&  & QR & 0.018 & 0.015 & 0.009 & 0.004 & 2.397 & 0.733 \\ 
		&  & IPW estimator & 0.020 & 0.016 & 0.010 & 0.007 & 2.879 & 0.662 \\ 
		&  & Weighted QR & 0.020 & 0.016 & 0.010 & 0.007 & 2.869 & 0.669 \\ 
		&  & G-comp (MC) & 0.014 & 0.012 & 0.007 & 0.002 & 2.289 & 0.745 \\ 
		&  & G-comp (approx) & 0.014 & 0.012 & 0.007 & 0.002 & 2.291 & 0.745 \\ \hline		
		
		Strong & 3 & Unadjusted & 0.023 & 0.015 & 0.009 & 0.008 & 3.106 & 0.796 \\ 
		&  & QR & 0.023 & 0.015 & 0.009 & 0.007 & 2.734 & 0.733 \\ 
		&  & IPW estimator & 0.027 & 0.018 & 0.010 & 0.017 & 4.559 & 0.662 \\ 
		&  & Weighted QR & 0.027 & 0.017 & 0.010 & 0.017 & 4.532 & 0.669 \\ 
		&  & G-comp (MC) & 0.018 & 0.012 & 0.007 & 0.003 & 2.476 & 0.745 \\ 
		&  & G-comp (approx) & 0.018 & 0.012 & 0.007 & 0.003 & 2.471 & 0.745 \\ \hline		
		
		Strong & 4 & Unadjusted & 0.032 & 0.015 & 0.009 & 0.020 & 5.633 & 0.796 \\ 
		&  & QR & 0.032 & 0.015 & 0.009 & 0.014 & 4.172 & 0.733 \\ 
		&  & IPW estimator & 0.038 & 0.018 & 0.010 & 0.029 & 6.956 & 0.662 \\ 
		&  & Weighted QR & 0.037 & 0.018 & 0.010 & 0.028 & 6.831 & 0.669 \\ 
		&  & G-comp (MC) & 0.024 & 0.011 & 0.007 & 0.008 & 3.304 & 0.745 \\ 
		&  & G-comp (approx) & 0.024 & 0.011 & 0.007 & 0.006 & 2.915 & 0.745 \\ 		
		\bottomrule
	\end{tabular}
	\endgroup
\end{table}

\section*{S5 \ \ Implementation of confounding-adjustment methods in LSAC example}

For the multivariable quantile regression approach, the outcome model included $A$ and $\bm{C}$ (consisting of 9 confounders; Table \ref{tab:lsac_variables}) as predictors with main effect terms only. The propensity score model (used for the IPW estimator and weighted quantile regression) regressed $A$ on $\bm{C}$ including main effect terms only. For both g-comp (MC) and g-comp (approx), the outcome model was specified as $\mathrm{log}(Y)$ conditional on $A$ and $\bm{C}$, including two-way interaction terms between $A$ and three confounders (child's sex, maternal completion of high school, consistent parenting) deemed plausible based on substantive knowledge. For g-comp (MC), we performed $R=1000$ draws per observation. For g-comp (approx), the vector of candidate $y^*$ values ranged over the values [0.01, 18] in increments of 0.01.

\clearpage
\newpage




\end{document}